\documentclass[aps,prl,twocolumn,showpacs,superscriptaddress,groupedaddress]{revtex4}  
\usepackage{graphicx}  
\usepackage{dcolumn}   
\usepackage{bm}        
\usepackage{amssymb}   
\usepackage{color}

\begin{document}

\leftline{Published in Physical Review Letters, 8 October 2014}


\title{Formation of Hard Power-laws in the Energetic Particle Spectra Resulting from Relativistic Magnetic Reconnection}
\author{Fan Guo}
\affiliation{Los Alamos National Laboratory, NM 87545 USA}
\author{Hui Li}
\affiliation{Los Alamos National Laboratory, NM 87545 USA}
\author{William Daughton}
\affiliation{Los Alamos National Laboratory, NM 87545 USA}
\author{Yi-Hsin Liu}
\affiliation{Los Alamos National Laboratory, NM 87545 USA}

\begin{abstract}

\noindent Using fully kinetic simulations, we demonstrate that magnetic reconnection in relativistic plasmas is highly efficient at accelerating particles through a first-order Fermi process resulting from the curvature drift of particles in the direction of the electric field induced by the relativistic flows.  This mechanism gives rise to the formation of hard power-law spectra in parameter regimes where the energy density in the reconnecting field exceeds the rest mass energy density $\sigma \equiv B^2/(4 \pi n m_ec^2) > 1$ and when the system size is sufficiently large.  In the limit $\sigma \gg 1$, the spectral index approaches $p=1$ and most of the available energy is converted into non-thermal particles.   A simple analytic model is proposed which explains these key features and predicts a general condition under which hard power-law spectra will be generated from magnetic reconnection.
\\

\end{abstract}

\maketitle


\textit{Introduction} -- Magnetic reconnection is a fundamental plasma process that allows 
rapid changes 
of magnetic field topology and the conversion of magnetic energy into plasma kinetic 
energy. It has been extensively discussed in solar flares, Earth's
magnetosphere, and laboratory applications. However,  
magnetic reconnection remains poorly understood in high-energy 
astrophysical systems
\citep{Hoshino2012}. 
Magnetic reconnection has been suggested as a mechanism for 
producing high-energy emissions 
from pulsar wind nebula, gamma-ray bursts, and jets from active 
galactic nuclei
\citep{Giannios2010,Kirk2004,Thompson1994,Zhang2011,Arons2012}. 
In those systems, it is often expected that the magnetization 
parameter $\sigma \equiv B^2/(4\pi n m c^2)$ exceeds unity.
Most previous kinetic studies focused on the non-relativistic regime $\sigma < 1$
 and reported several acceleration
mechanisms such as acceleration at X-line regions \citep{Pritchett2006,Fu2006,Oka2010} and Fermi-type
acceleration within magnetic islands \citep{Pino2005,Drake2006,Fu2006,Oka2010}.
More recently, the regime $\sigma= 1$-$100$ has
been explored using pressure-balanced
current sheets and strong particle acceleration has been found in both diffusion regions 
\citep{Zenitani2001,Zenitani2007,Cerutti2013,Sironi2014} and island regions 
\citep{Liu2011,Bessho2012}. However, this initial 
condition requires a hot plasma component inside the current sheet to maintain 
force balance, which may not be justified for high-$\sigma$ plasmas.

For magnetically dominated systems, it has been shown \citep{Titov2003,Galsgaard2003}  that the gradual evolution of the magnetic field can lead to formation of intense nearly force-free current layers where magnetic reconnection may be triggered. In this Letter, we 
perform large-scale two-dimensional (2D) and three-dimensional (3D) full 
particle-in-cell (PIC) simulations of 
a relativistic force-free  
current sheet
with $\sigma$ up to $1600$. 
In the high-$\sigma$ regime, the release of 
magnetic energy is accompanied by 
the energization of nonthermal particles on the same fast time scale as the reconnection process. Much of the 
magnetic energy is converted into the kinetic energy of
nonthermal relativistic particles and the
eventual energy spectra show
a power-law $f(\gamma) \propto \gamma^{-p}$ over nearly two
decades, with the spectral index $p$ decreasing
with $\sigma$ and system size, and approaching 
$p=1$. The dominant acceleration mechanism is a first-order Fermi process 
through the curvature drift motion of particles
along the electric field
induced by relativistic
reconnection outflows. The formation of the power-law distribution can be 
described by a simple model that includes both inflow and the Fermi 
acceleration. This model also appears to explain recent PIC 
simulations \citep{Sironi2014}, which reported hard power-law 
distributions after subtracting the initial hot plasma 
population inside the current layer.

\begin{figure}
\includegraphics[width=0.5\textwidth]{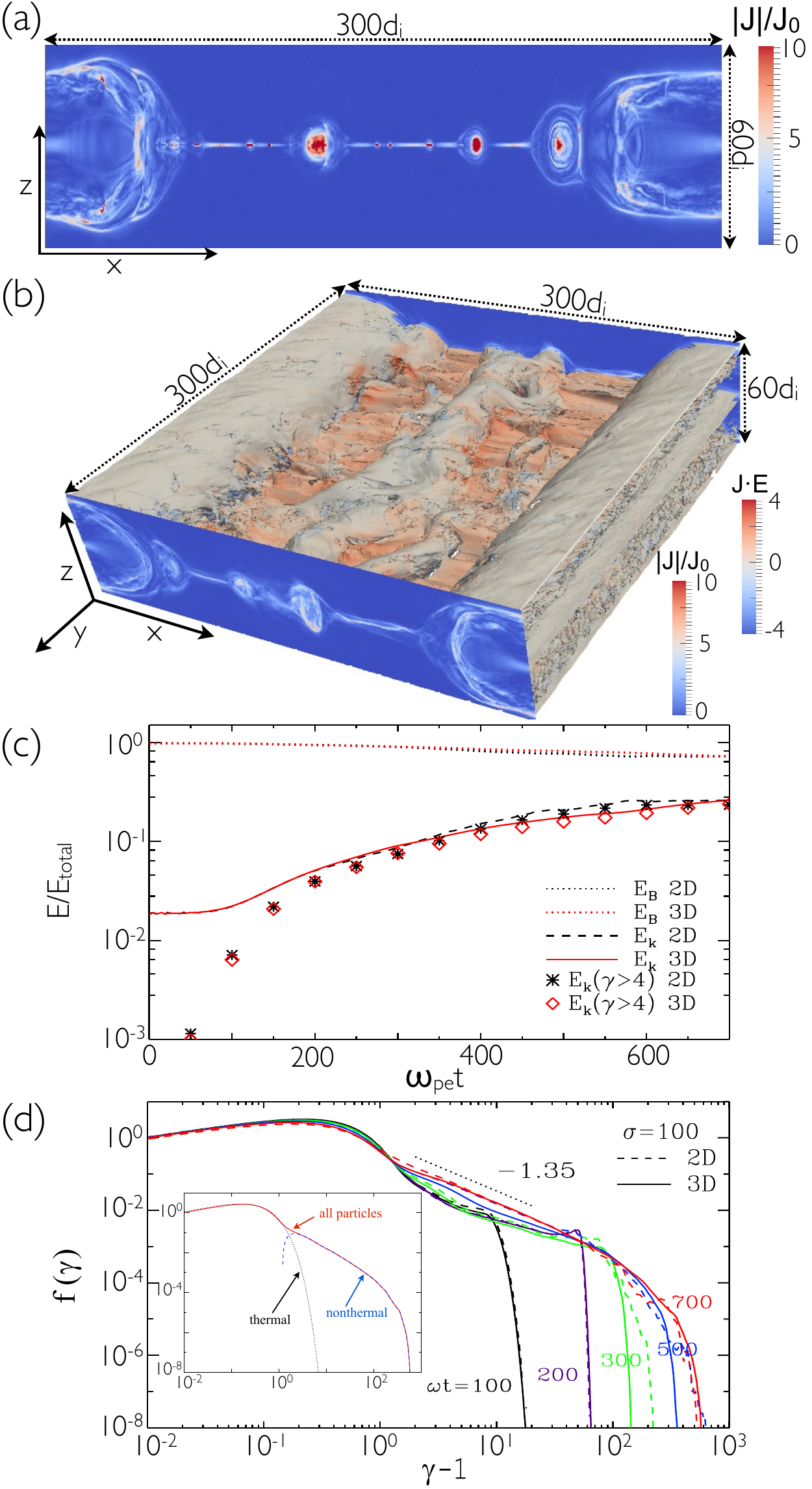}
\caption{\label{fig:epsart} Results from 2D and 3D PIC simulations with $\sigma = 100$. (a) Current density from 2D simulation at $\omega_{pe}t=375$. (b) $x$-$z$ cut of current density and an isosurface of current density with color-coded $\textbf{J}\cdot \textbf{E}$ normalized using $n_0m_ec^2 \omega_{pe}$ at $\omega_{pe}t=375$. (c) Evolution of magnetic energy $E_B$, total kinetic energy $E_k$, and kinetic energy carried by relativistic particles with $\gamma >4$. (d) Evolution of particle energy spectra from 2D and 3D simulations. Subpanel:  energy spectrum from the 3D simulation at $\omega_{pe}t=700$. }
\end{figure}

\textit{Numerical simulations} -- The initial condition is a 
force-free current layer with $\textbf{B} = B_0 \text{tanh} 
(z/\lambda) \hat{x} + B_0 \text{sech} (z/\lambda) \hat{y}$, which 
corresponds to a magnetic field with magnitude $B_0$ 
rotating by $180^\circ$ across the layer with
a thickness of $2\lambda$. The plasma consists of 
electron-positron pairs with mass ratio $m_i/m_e = 1$. The initial distributions 
are Maxwellian with a uniform density $n_0$ 
and temperature ($T_{i}=T_{e}=0.36 m_ec^2$). 
Particles in the sheet have a net drift $\textbf{U}_i 
= - \textbf{U}_e$ to give a current density $\textbf{J} = en_0(\textbf{U}_i - \textbf{U}_e)$ consistent with 
$\nabla \times \textbf{B} = 4\pi \textbf{J}/c$. The simulations are 
performed using the VPIC \citep{Bowers2009} and NPIC codes \citep{Daughton2006,Daughton2007}, both of 
which solve the relativistic Vlasov-Maxwell system of equations. 

In the simulations, $\sigma$ is adjusted by 
changing the ratio of the electron gyrofrequency to 
plasma frequency $\sigma = B^2/(4\pi n_e m_e c^2) = (\Omega_{ce}/
\omega_{pe})^2$.
A series of 2D simulations were performed with
$\sigma = 1 \rightarrow 1600$ and domain sizes 
$L_x\times L_z =300 d_i \times 
194 d_i$, $600 d_i \times 388 d_i$, and $1200 d_i 
\times 776 d_i$, where $d_i = c/\omega_{pe}$ is the inertial length. For 3D simulations, the largest case 
is $L_x \times L_y \times L_z = 300 d_i \times 300 
d_i \times 194 d_i$ with 
$\sigma=100$. For high-$\sigma$ cases, we choose 
grid sizes $\Delta x = \Delta y = 
1.46/\sqrt{\sigma} d_i$ and $\Delta z = 0.95/\sqrt{\sigma} d_i$, so 
the gyroradius $\sim v_{the} d_i/(\sqrt{\sigma}c)$ is resolved. The half-thickness is $\lambda = 6 d_i$ for $\sigma \leq 100$, $12 d_i$ for $\sigma=400$, and $24 d_i$ for $\sigma=1600$ in order to satisfy $\text{U}_i < c$.
All simulations used more than 100 particles per cell for each species, employed periodic boundary conditions in the $x$- and $y$-directions, and in the $z$-direction used conducting boundaries for the fields and reflecting for the particles. A long-wavelength perturbation \citep{Daughton2007} with $B_z = 0.03 B_0$ is included to 
initiate reconnection.


\textit{Simulation results} -- Figure 1 contrasts some key results from 2D and 3D simulations with $\sigma=100$ and domain size $L_x \times L_z = 300d_i \times 194 d_i$ ($L_y=300d_i$ for the 3D simulation). Panel (a) 
shows the current density at $\omega_{pe}t=375$ in the 2D simulation. Because of the secondary tearing instability, several fast-moving secondary plasmoids develop along the central region and merge to form larger plasmoids \citep{Daughton2007}.
Panel (b) shows an isosurface of current 
density colored by $\textbf{J}\cdot 
\textbf{E}$ at $\omega_{pe}t=375$ from the 3D simulation. As the initial guide field is expelled outward from the central region, the kink instability \citep{Daughton1999} develops and interacts with the tearing mode, leading to a turbulent evolution \citep{Daughton2011}. Previous studies have suggested different predictions concerning the influence of $\sigma$ on the reconnection rate \citep{Blackman1994,Lyutikov2003,Lyubarsky2005,Zenitani2009,Melzani2014}. In this letter, the reconnection rate is observed to increase with $\sigma$ from $E_{rec} \sim 0.03B_0$ for $\sigma=1$ to $E_{rec} \sim 0.22B_0$ for $\sigma=1600$.
Although the 2D and 3D 
simulations appear quite different, the energy conversion and 
particle energization are surprisingly similar.
Panel (c) compares the evolution of magnetic energy $E_B$, plasma 
kinetic energy $E_k$, and energy in relativistic particles with 
 $\gamma > 4$. In both cases, about $20\%$ of the 
magnetic energy is converted into kinetic energy of relativistic particles. Figure 1 (d) 
compares the energy spectra at various 
times. The most striking feature is that a hard power-law 
spectrum with index $p \sim 1.35$ forms in both 2D 
and 3D runs. In the subpanel, the energy spectrum for all 
particles in the 3D simulation at $\omega_{pe}t=700$ is shown 
by the red line. The low-energy portion can be fitted by a Maxwellian 
distribution (black) and the nonthermal part 
resembles a power-law distribution (blue) starting at $\gamma \sim 2$ with an exponential cut-off apparent for $\gamma \gtrsim 100$. The nonthermal 
part contains $\sim25\%$ of particles and $\sim95\%$ of the
kinetic energy. The maximum particle energy is predicted  approximately using the
reconnecting electric field 
$m_ec^2\gamma_{max} = \int |qE_{rec}|c dt$ until 
the gyroradius is comparable to the system size. 
Although we observe a strong kink instability in the 3D simulations, the energy conversion and particle 
energy spectra are remarkably similar to the 2D results, indicating the 3D effects 
are not crucial for understanding the particle acceleration. Since there is more
freedom to vary the parameters in 2D simulations, in the rest of this 
letter we focus on this limit.

In Figure 2, we present more analysis for the acceleration mechanism using the case with 
$\sigma=100$ and $L_x\times L_z=600d_i \times 388 d_i$. Panel (a) shows the energy as a 
function of the $x$-position of four accelerated particles.  The electrons gain energy by bouncing back and forth within the reconnection layer. 
Upon each cycle, the energy gain is $\Delta \gamma \sim \gamma$, which demonstrates that 
the acceleration mechanism is a first-order Fermi process 
\citep{Drake2006,Drake2010}. To show this more rigorously, we have 
tracked the energy change 
of all the particles in the simulation and contributions
from the parallel electric field 
($m_ec^2\Delta \gamma = \int qv_{\parallel} E_\parallel dt$) 
and curvature drift 
acceleration ($m_ec^2 \Delta \gamma = \int q \textbf{v}_{curv} \cdot \textbf{E}_{\perp} dt$) similar to \citep{Dahlin2014}, 
where $\textbf{v}_{curv} = \gamma v^2_\parallel (\textbf{b} \times (\textbf{b}\cdot \nabla)\textbf{b})/\Omega_{ce}$, $v_\parallel$ is the particle velocity parallel to the magnetic field, and $\textbf{b} = \textbf{B} / |B|$.
Panel (b) shows the averaged energy gain and the contribution from parallel 
electric field and curvature drift acceleration over an 
interval of $25 \omega_{pe}^{-1}$ as a function of energy starting at $\omega_{pe}t=350$. 
The energy gain follows $\Delta \gamma \sim \alpha \gamma$, confirming the first-order 
Fermi process identified from particle trajectories. The energy gain 
from the parallel motion is weakly dependent on energy, whereas the energy gain from the
curvature drift acceleration is roughly proportional to energy. In the early phase, the parallel electric field is strong but only 
accelerates a small portion of particles, and the curvature drift  dominates the 
acceleration starting at about $\omega_{pe}t = 250$. 
The contribution from the gradient drift was also evaluated and found to be unimportant. Panel (c) shows  $\alpha = 
<\Delta \gamma>/(\gamma \Delta t)$ measured directly from the energy gain of the particles in the perpendicular electric field ($m_ec^2 \Delta 
\gamma = \int q \textbf{v}_{\perp} \cdot \textbf{E}_{\perp} dt$) and estimated from the 
expression for the curvature drift acceleration.
The close agreement demonstrates that curvature drift 
term dominates the particle energization. For higher $\sigma$ and 
larger domains, the acceleration is stronger and reconnection is sustained over a longer 
duration. In panel (d), a 
summary for the observed spectral index of all the 2D runs shows that the spectrum is harder for 
higher $\sigma$ and larger domain sizes, and approaches the limit $p=1$. 

\begin{figure}
\includegraphics[width=0.5\textwidth]{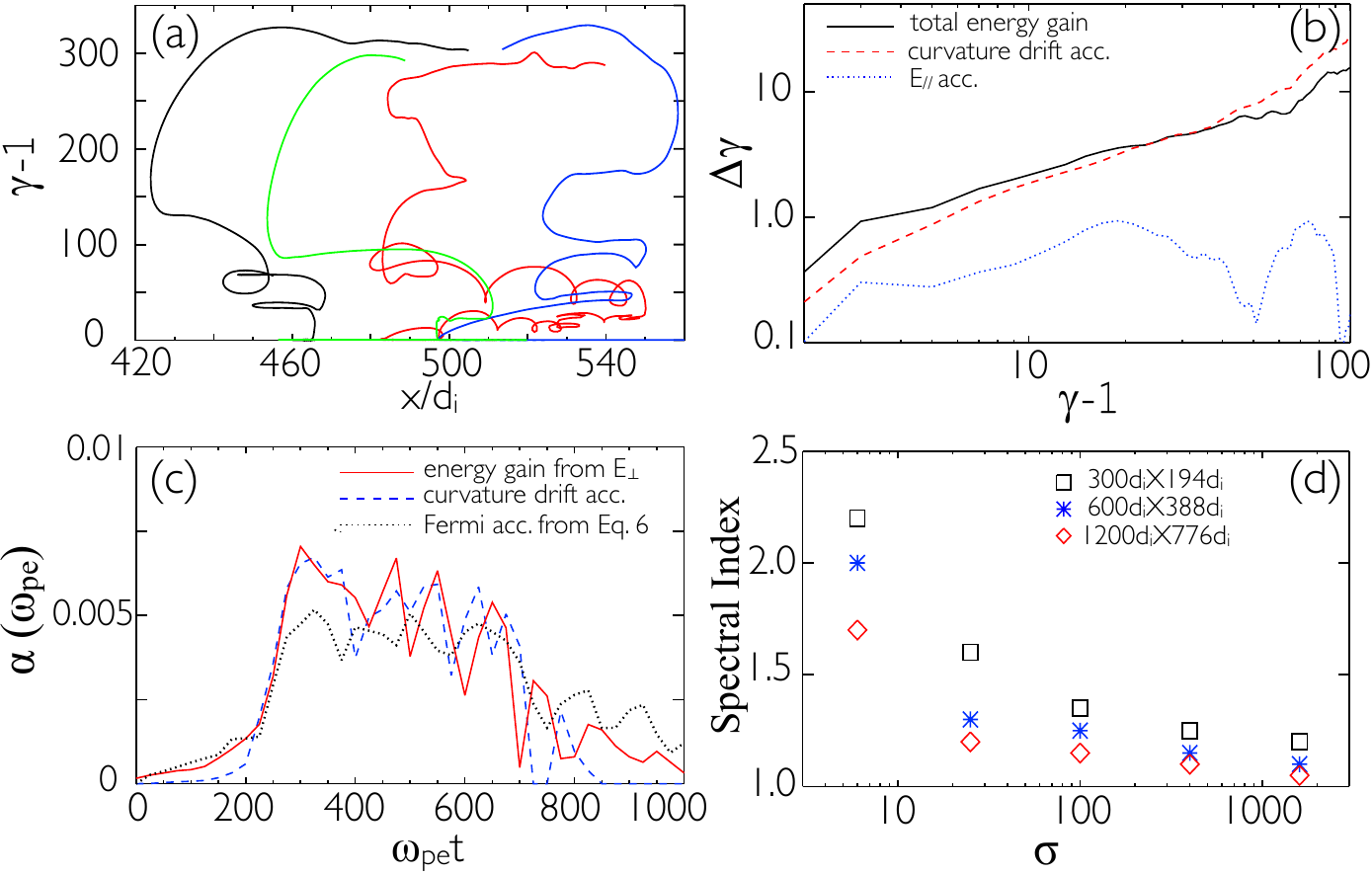}
\caption{(a) Energy as a function of $x$-position of four accelerated particles; (b) 
Averaged energy gain and contributions from parallel electric fields and curvature drift 
acceleration over an interval of $25 \omega_{pe}^{-1}$ as a function of particle energy 
starting at $\omega_{pe}t=350$; (c) $\alpha = <\Delta \gamma>/(\gamma \Delta t)$ from 
energy gain in perpendicular electric field and by curvature drift acceleration, and from 
the Equation (6) using the averaged flow speed and island size; (d) Spectral index of all 
2D simulations.}
\end{figure}

\textit{New Model} -- It is often argued that some loss mechanism 
is needed to form a power-law distribution 
\citep{Zenitani2001,Drake2010}. However, the simulation results reported here illustrate clear power-law 
distributions 
in a closed system. Here we demonstrate that these 
results can be understood in terms of a model 
illustrated in Figure 3 (a). As reconnection proceeds, 
the ambient plasma is injected into the 
acceleration region at a speed $V_{in} = c\textbf{E}_{rec}\times \textbf{B} /B^2$. We consider the continuity equation for the energy distribution function $f(\varepsilon, t)$ within the acceleration region
\begin{eqnarray}
\frac{\partial f}{\partial t} + \frac{\partial}{\partial \varepsilon} \left( \frac{\partial \varepsilon}{\partial t} f \right) = \frac{f_{inj}}{\tau_{inj}} - \frac{f}{\tau_{esc}},
\end{eqnarray}

\noindent with $\partial \varepsilon/\partial t = \alpha \varepsilon$, where $\alpha$ is 
the constant acceleration rate from the first-order Fermi process, 
$\varepsilon = m_e c^2 (\gamma-1)/T$ is the normalized kinetic energy, $\tau_{inj}$ is 
the time scale for injection of particles from the upstream region with fixed distribution
$f_{inj}$ and $\tau_{esc}$ is escape time.  We assume that the initial distribution within the layer $f_0$ and the upstream injected distribution are both Maxwellian with initial temperature $T < m_e c^2$ such that 
\begin{eqnarray}
f_{inj} &\propto & \gamma (\gamma^2-1)^{1/2} \exp (-\varepsilon) \\ \nonumber
        &\approx & \sqrt{2 \varepsilon} \left(1 + \frac{5T}{4 m_e c^2} \varepsilon + .... \right) \exp(-\varepsilon) \;.
\end{eqnarray}
For simplicity, we consider the lowest order (non-relativistic) term in this expansion and normalize $f_0 
=\frac{2 N_0}{\sqrt{\pi}}\sqrt{\varepsilon}\exp(-\varepsilon)$ by the number of particles $N_0$ within the initial layer and $f_{inj}$ by the number of particles injected into the layer $N_{inj} \propto V_{in} \tau_{inj}$ during reconnection. With these assumptions,  the solution to (1)  can be written as
\begin{eqnarray} \label{time-solution}
 f(\varepsilon,t) &=& \frac{2 N_0}{\sqrt{\pi}} \sqrt{\varepsilon} e^{-(3/2+\beta ) \alpha t} \exp(-\varepsilon e^{-\alpha t}) \\ \nonumber
 &+& \frac{2 N_{inj}}{\sqrt{\pi}(\alpha \tau_{inj}) \varepsilon^{1+\beta}} \left[ 
 \Gamma_{(3/2+\beta)}(\varepsilon e^{-\alpha t}) - \Gamma_{(3/2+\beta)}(\varepsilon) \right] ,
\end{eqnarray} 
\noindent where $\beta = 1/(\alpha \tau_{esc})$ and $\Gamma_s (x)$ is the 
incomplete Gamma function. The first term accounts for particles initially in the acceleration region while the 
second term describes the evolution of injected particles. In the limit of no injection or 
escape ($\tau_{esc} \to \infty$ and $\tau_{inj} \to \infty$), the first term in (3) remains a thermal distribution with enhanced temperature $e^{\alpha t}T$, consistent with
Ref. \citep{Drake2010}. However, as reconnection proceeds new particles 
enter continuously into the acceleration region and due to the periodic boundary conditions there is no particle escape. Thus considering the case $\tau_{esc} \to \infty$ and assuming $N_0 \ll N_{inj}$, at the time $t=\tau_{inj}$ when reconnection saturates the second term in (3) simplifies to 
\begin{eqnarray}
f(\varepsilon, \tau_{inj}) = \frac{N_{inj}}{\alpha \tau_{inj}} [ \frac{ \text{erf}(\varepsilon^{1/2}) - \text{erf}(\varepsilon^{1/2}e^{-\alpha \tau_{inj}/2})}{\varepsilon} \\ \nonumber
 + \frac{2}{\sqrt{\pi}} \frac{e^{-\alpha \tau_{inj} /2} \exp(-\varepsilon e^{-\alpha \tau_{inj}}) - e^{-\varepsilon}}{\varepsilon^{1/2}} ].      %
\end{eqnarray}


\noindent When $\alpha \tau_{inj} > 1$, this gives the relation $f \propto 1/\varepsilon$ 
in the energy range $1 < \varepsilon < e^{\alpha \tau_{inj}}$ as shown in Figure 3 (b) by 
directly evaluating (4) for different $\alpha \tau_{inj}$. Interestingly, 
this energy range for the power-law is below that of the heated thermal particles in the 
initial layer.  Thus in the limit $N_0 \sim N_{inj}$ the first term in (3) should be 
retained and the power-law produced is sub-thermal relative to this population. While it is 
straightforward to obtain the relativistic corrections arising from the injected 
distribution (2), we emphasize that these terms do not alter the spectral 
index.


In order to estimate the acceleration rate $\alpha$, the energy change of each particle can be approximated by a relativistic collision formula \citep[e.g.,][]{Longair1994}
\begin{eqnarray}
\Delta \gamma =\left(\Gamma_V^2(1+\frac{2Vv_x}{c^2}+\frac{V^2}{c^2}) - 1 \right)\gamma,
\end{eqnarray}

\noindent where $V$ is the outflow speed, $\Gamma_V^2 = 1/(1-V^2/c^2)$, and $v_x$ is the particle velocity in the $x$-direction. The time between two collisions is about $L_{is}/v_x$, where $L_{is}$ is the typical size of the magnetic islands (or flux ropes in 3D). Assuming relativistic particles have a nearly isotropic distribution $v_x \sim c/2$
\begin{eqnarray}
\alpha \sim \frac{c(\Gamma_V^2(1+\frac{V}{c}+\frac{V^2}{c^2})-1)}{2L_{is}}.
\end{eqnarray}
\noindent Using this expression, we measure the averaged $V$ and $L_{is}$ from the
simulations and estimate the time-dependent acceleration rate $\alpha (t)$. An example is shown in Figure 2 (c). This agrees reasonably well with that obtained from perpendicular acceleration and curvature drift acceleration. 
Figure 3 (c) shows the time-integrated value of $\alpha \tau_{inj} = \int^{\tau_{inj}}_0 \alpha(t) dt$ for various simulations with $\sigma = 6 - 400$.  For cases with $\alpha \tau_{inj} > 1$, a hard power-law distribution with spectral index $p \sim 1$ forms. 
For higher $\sigma$ and larger system size, the magnitude of $\alpha \tau_{inj}$ 
increases approximately as $\propto \sigma^{1/2}$. 

\textit{Discussion} -- Considering the more realistic limit with both particle loss and injection, Equation (3) predicts a spectral index $p = 1 + 1/(\alpha \tau_{esc})$ when $\alpha \tau_{inj} > 1$, recovering the classical Fermi solution 
\citep[e.g.,][]{Longair1994}. If the escape is caused by convection out of the acceleration region $\tau_{esc} = L_x/V$, the spectral index should approach  $p = 1$ when $\alpha \tau_{esc} \gg 1$ in the high-$\sigma$ regime. Although the present simulations employed periodic boundary conditions, most cases develop power-law distributions within two light-crossing times, indicating that the boundary conditions do not strongly influence 
the results. In preliminary 2D simulations using open boundary conditions \citep{Daughton2006}, we have confirmed these general trends [Guo et al. 2014, in preparation].
For non-relativistic reconnection, the acceleration rate is
lower and thus it takes longer to form a power-law distribution. Take the nonrelativistic 
limit for (6), if $V = 0.1c$, $v_x=0.2c$, and $L_{is} = 100 
d_i$, the reconnection has to be sustained over a time $\tau_{inj} > 2 \times 10^4 
\omega_{pi}^{-1}$ to form a power law, which significantly 
exceeds the simulation time of most previous studies. It has been 
suggested that current sheet instabilities may strongly 
influence particle acceleration \citep{Zenitani2007}. In
contrast, the energy distributions reported here are remarkably similar in 2D 
and 3D, despite the broad range of secondary kink and tearing instabilities in 3D.   This surprising result suggests that the 
underlying Fermi acceleration is rather robust and does not depend on the
existence of well-defined magnetic islands. The strong similarities between the 2D and 3D 
acceleration spectra are also consistent with some key similarities in the reconnection 
dynamics.  In particular, the range of scales for the 2D magnetic islands is similar to the 
observed 3D flux ropes.  In addition, the reconnection rate and flow speeds are also 
quite similar between 2D and 3D, in agreement with other recent studies \citep{Liu2013,Daughton2014}.
In large open systems, it remains to be seen whether 3D turbulence may affect the particle 
escape times. Another important factor that may influence these results is the presence of 
an external guide field $B_g$.  Our preliminary simulations suggest that the key results 
of this letter will hold for $B_g < B_0$. 
 For stronger guide fields, the energy release is slower and the associated particle 
 acceleration requires further study.

\begin{figure}
\includegraphics[width=0.5\textwidth]{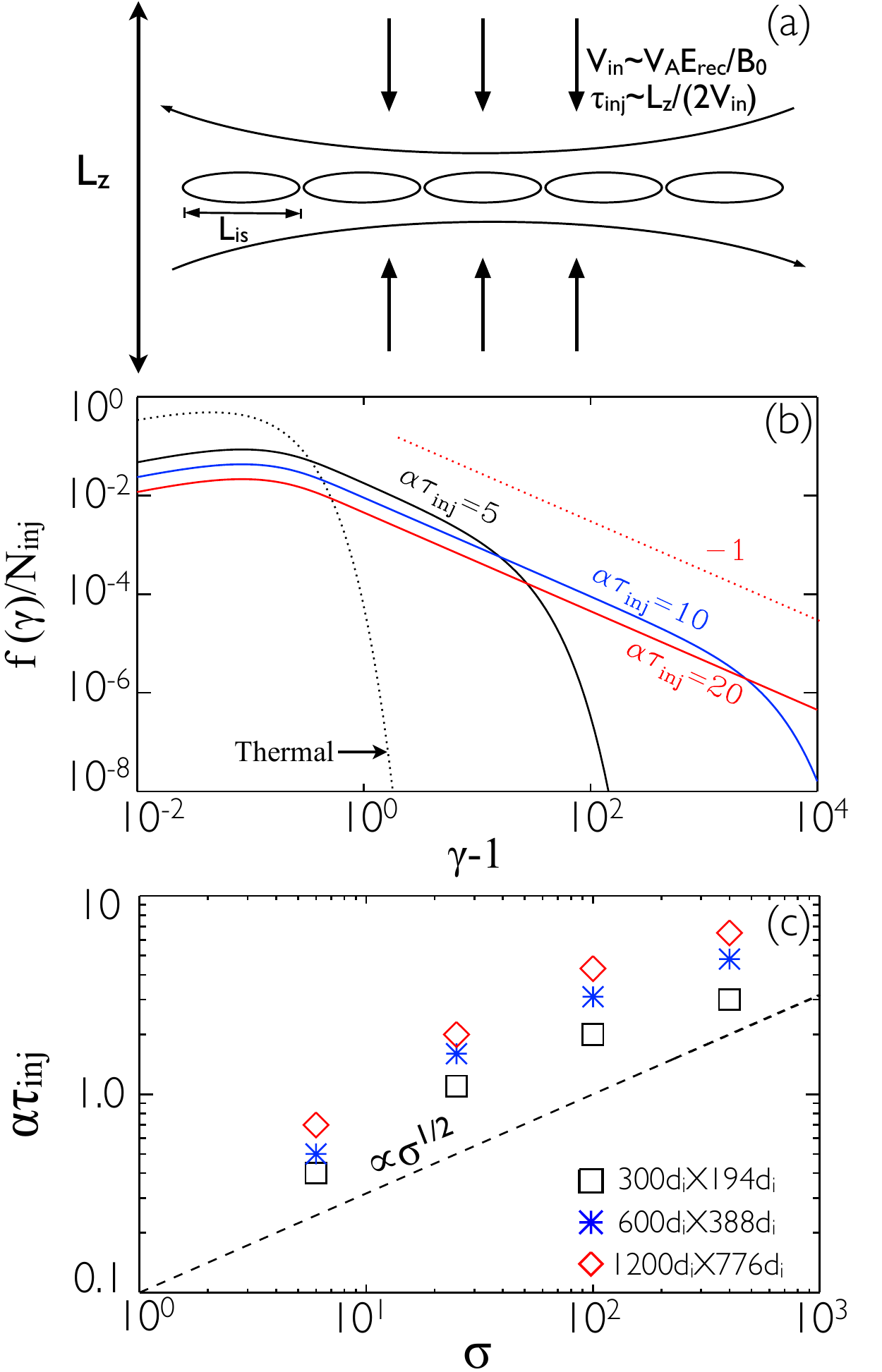}
\caption{\label{fig:epsart4} (a) Illustration of the acceleration model for the formation of power law distributions; (b) Analytical results for different $\alpha \tau_{inj}$ obtained from (4); (c) Time integrated $\alpha \tau_{inj}$ for cases with $\sigma = 6-400$ and different system sizes.}
\end{figure}

We have demonstrated that in the regime 
$\sigma \gtrsim 1$ magnetic reconnection is an 
efficient mechanism of converting the energy stored in 
the magnetic shear into relativistic 
nonthermal particles. These energetic particles contain a significant fraction of the total energy released and, quite 
interestingly, have a power-law energy distribution with 
spectral index $p \sim 1$ when $\alpha \tau_{inj} > 1$. Physically, this requires that the time scale over which particles are injected into the acceleration region is longer than acceleration time for the first-order Fermi process.    The results in this letter demonstrate this condition is more easily achieved in regimes with $\sigma \gg 1$, but may also occur with $\sigma \gtrsim 1$ in sufficiently large reconnection layers. 
Our new findings substantiate the importance of fast magnetic reconnection in 
strongly magnetized plasmas, and may be important for explaining the 
high-energy emissions in systems like pulsars, jets from black 
holes, and gamma-ray bursts.

\textit{Acknowledgements} -- We thank the referees for providing helpful reports for improvement 
and clarification to the paper, especially the insightful suggestions regarding the 
analytical derivation.
We gratefully acknowledge discussions 
with Andrey Beresnyak and Dmitri Uzdensky. We are grateful for support from DOE 
through the LDRD program at LANL and DoE/OFES support to LANL in 
collaboration with CMSO.  This research is part of the Blue 
Waters sustained-petascale computing project, which is supported 
by the NSF (OCI 07-25070) and the state of Illinois. Additional 
simulations were performed at the National Center for 
Computational Sciences at ORNL and with LANL institutional 
computing.

\end{document}